%% file: main.tex
\documentclass[a4paper, amsfonts, amssymb, amsmath, preprint, showkeys, nofootinbib, twoside]{revtex4-1}
\usepackage[english]{babel}
\usepackage[utf8]{inputenc}
\usepackage[colorinlistoftodos, color=green!40, prependcaption]{todonotes}
\usepackage{soul}
\input{preamble}
\usepackage[pdftex, pdftitle={Article}, pdfauthor={Author}]{hyperref} 
\usepackage{algorithm}
\usepackage{algpseudocode}
\begin{document}

\title{High-energy picosecond pulses with a single spatial mode from a passively mode-locked, broad-area semiconductor laser}
\author{Mallachi-Elia Meller, Leon Bello, Idan Parshani, Yosef London, Avi Pe'er}
    \email[Correspondence email address: ]{avi.peer@biu.ac.il}
     \affiliation{Department of Physics and BINA Institute of Nanotechnology, Bar-Ilan University, Ramat-Gan 52900, Israel}
    
\date{\today} 

\begin{abstract}
We present a mode-locked semiconductor laser oscillator that emits few picosecond pulses (5-8ps at 379MHz repetition) with record peak power (112W) and pulse energy (0.5nJ) directly out of the oscillator (with no amplifier). To achieve this high power performance we employ a high-current broad-area, spatially multi-mode diode amplifier (0.3×5mm), placed in an external cavity that enforces oscillation in a single spatial mode. Consequently, the brightness of the beam is near-ideal ($M^2 = 1.3$). Mode locking is achieved by dividing the large diode chip (edge emitter) into two sections with independent electrical control: one large section for gain and another small section for a saturable absorber. Precise tuning of the reverse voltage on the absorber section allows to tune the saturation level and recovery time of the absorber, which provides a convenient control knob to optimize the mode-locking performance for various cavity conditions. 
\end{abstract}


\maketitle
 
\section{Introduction}
Semiconductor lasers represent the industry standard for sources of coherent light due to their high electrical-to-optical efficiency, simple and robust construction, high optical power, tunability, wide range of available wavelengths (covering the entire VIS-NIR range), and low cost. Thus, the generation of ultrashort optical pulses directly from semiconductor lasers with high peak power and high spatial beam quality (brightness) is highly desired for many applications, such as: 3D micromachining of polymer materials  \cite{application_tpa1,application_tpa2,application_tpa3}, dvanced methods of nonlinear and fluorescence microscopy \cite{microscopy1,microscopy2,microscopy3,microscopy4}, precision measurement and frequency metrology \cite{metrology1,metrology2,metrology3}, frequency conversion \cite{applications3,applications4}, direct material processing \cite{applications5,applications6} and medical applications \cite{applications7}. For these applications however, the peak power of laser diodes oscillators so far is not up to par with standard solid-state or fiber lasers, such as mode-locked Ti:Sapphire lasers or Mode locked Er / Yt fiber lasers, which offer the more prevalent solution in the market currently, despite their high complexity, limited wavelength availability and higher cost.

Since the optical power of a laser diode is directly dictated by its physical area, broad area laser diodes (BAL) are an attractive route for high-power lasers competitive with solid-state lasers, while maintaining high customizability and simplicity -- requiring only electrical pumping. Unfortunately, this high power of BAL diodes normally comes at the expense of degraded spatial coherence since the wide cross-section of the diode wave-guide is inherently spatially multi-mode along the slow axis.

To mitigate the degradation of brightness, many techniques were developed, ranging from shaping the spatial profile of diode laser chip itself by special fabrication techniques \cite{Single-Mode_3, Single-Mode_4} to configurations of coherent beam combining \cite{CBC_1, CBC_2, CBC_3}. However, the simplest approach that maintains the simplicity of the multimode BAL chip, is placing the BAL gain medium in an external cavity \cite{ex_cavity_1, ex_cavity_2,prev_mallachi}, which shapes the spatio-temporal profile of the beam and enforces single-mode operation. The external cavity approach is advantageous for a couple of reasons -- it is simple, it allows great flexibility, and requires only off-the-shelf components. However, high-power operation is difficult, mainly due to self-lasing of the diode. At high powers, even small parasitic reflections are enough to induce self-lasing inside the diode, bypassing the external feedback. In this work, we circumvent this limitation using a simple and flexible design based on an angle-cut diode, which strongly suppresses self lasing, allowing us to obtain high-power mode locked operation, as reported hereon

In a previous paper \cite{prev_mallachi}, we have employed an external cavity to combine power and brightness from a single diode oscillator \textit{in continuous-wave operation}. Here we implement a similar concept with a passively mode locked semiconductor laser to obtain both high pulse-energy and spatial coherence in a single oscillator. We achieve record results for a single diode oscillator, with pulse-energy of ($0.55\rm{nJ}$) and peak power ($112 \rm{W}$), along with a short pulse duration ($4.9\rm{ps}$), all in a highly pure spatial mode ($M^2 = 1.3$). This is achieved in a simple design that uses no special fabrication techniques, only off-shelf components and no subsequent amplification stages. Of course, amplification and custom-made diodes are still compatible with our design, and could be used to improve our results even further. Mode locking is obtained using a diode chip with two-sections that are independently electrically controlled: a large section of the diode is pumped with forward current to act as the gain medium, and a smaller section of the diode that is reverse-biased to act as a saturable absorber with tunable saturation \cite{BOOK1,BOOK2,BOOK3}. We demonstrate mode-locking at even-harmonics operation \cite{harmonics_ML_1, harmonics_ML_2,harmonics_ML_3}, characteristic of colliding mode-locking \cite{CPM_1, CPM_2}, and characterize the spatial and temporal performance of the system for different harmonics.

\section{Experiment}
\begin{figure}[ht]
    \centering
    \includegraphics[width = 1\textwidth]{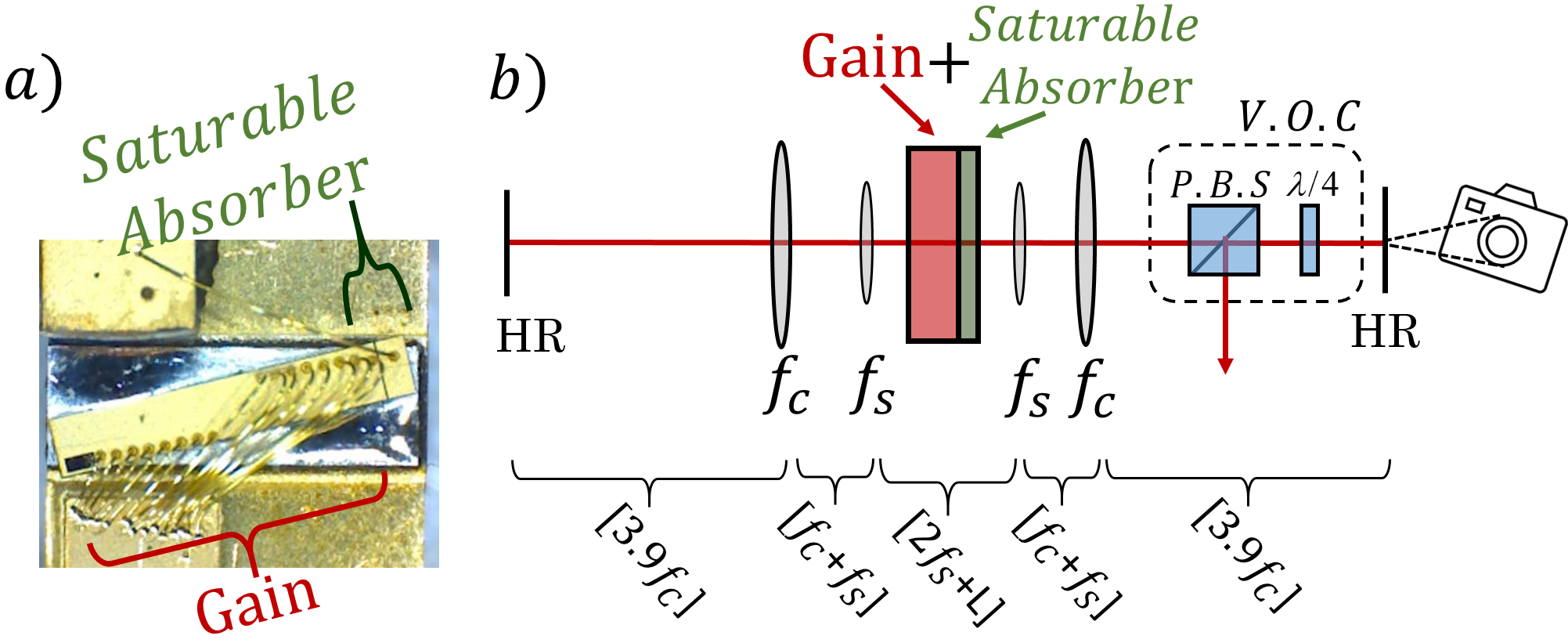}
    \caption{\textbf{(a)} Close up image of our angle-cut diode, with the two different sections highlighted. The longer section is operated with forward voltage and acts as a gain medium, while the shorter one is driven in reverse voltage and acts as a variable saturable absorber. \textbf{(b)} Simplified scheme of our external cavity configuration. Our configuration is comprised of a two-section angle-cut diode, separating the two arms of the cavity. Each arm includes a cylindrical lens and a spherical lens to stabilize the slow and fast axes of mode, while the left arm also includes a variable output coupler (denoted V.O.C in the figure).}
    \label{fig:scheme}
\end{figure}

Figure \ref{fig:scheme}a shows an image of the diode chip (5x0.3mm), which is cut at an angle to mitigate parasitic feedback from the diode facets that may cause self-lasing. The diode is comprised of a large gain section and the small saturable absorber section with independent electrical contacts, as marked on the image. The gain section is driven in forword current to provide amplification, whereas the absorber section is driven in reverse voltage to tune its saturation level to induce passive mode-locking and optimize its performance. This diode chip was then placed in the external cavity configuration of Figure \ref{fig:scheme}b, comprised of two identical cavity arms around the diode chip (of ~400mm length each). The length of each arm acts as a soft spatial filter \cite{prev_mallachi} that enforces oscillation in a single spatial mode. the left arm includes also a variable output coupler, implemented by a polarizing beam-splitter and a rotating $ \lambda/4$ wave-plate for optimizing the mode-locking performance and output power (see caption of Figure \ref{fig:scheme} for details). To stabilize the cavity for both the fast and slow axes, the diode facet is located slightly before the focal plane of a short spherical fast lens ($f_s=9mm$), thereby imaging the diode facet onto the end mirror of the cavity (at ~400mm distance). An additional cylindrical lens of a longer focus ($f_c=75mm$) is placed to form a telescope for the slow axis (together with the spherical lens). This arrangement stabilizes the slow-axis mode of the cavity and enforces single mode operation,

We evaluated the performance of our system both temporally and spatially. The temporal width of the ultrashort pulse was measured using an intensity autocorrelator (model FR-103XL with a resolution of $0.1 \rm{ps}$) and to evaluate the beam profile we employed a simple high-resolution CCD camera (Thorlabs model CS165MU with spatial resolution of $ 3.45\mu m$). We investigated the influence of the saturable absorber voltage on the the temporal profile of the generated pulses. In order to characterize the repetition-rate of our pulse-train and its stability, we used a fast photodetector followed by an RF spectrum analyser (Keysight model N9020A).

\section{Results}
Table \ref{tab:parameters} summarizes the measured performance of our laser for various harmonics of the fundamental repetition rate (2nd-to-8th with repetition rates of $380-1520\rm{MHz}$). Since pulses were formed only for even harmonics of the repetition rate, we suspect that the interaction between colliding pulses in the gain medium was involved in the mode-locking mechanism \cite{CPM_1, CPM_2}, but to verify this further studies will be needed. For the 2nd harmonic we obtained high pulse energy of $550 \rm {pJ}$ and $4.9 \rm ps$ pulse duration. The spatial purity of the beam was practically single-mode with $M^2=1.2-1.4$ and fractional power in the main spatial lobe of $86-88\%$ for all the harmonics.

\begin{table*}[ht]
\centering
    \begin{tabular}{| p{2.3cm} ||p{1.1cm}|p{1.1cm}|p{1.4cm}|p{1.5cm}|p{1cm}|p{1.3cm}|p{1.3cm}|}
         \hline
         \multicolumn{8}{|c|}{\textbf{Performance and parameters}} \\
         \hline
         \# Harmonic & Pulse energy [pJ] & Peak power [W] & Average power [mW] & Pulse duration [ps] & TBP & Current [A] & Reverse voltage [V] \\
         \hline
         \textbf{2$^{nd}$} & \textbf{551} & \textbf{112.6} & \textbf{209} & \textbf{4.9} & \textbf{5.7} & \textbf{5.19} & \textbf{4.2} \\
         \hline 
         \textbf{4$^{th}$} & 366 & 69.9 & 278 & 5.2 & 3.5 & 5.21 & 3.6 \\
         \hline 
         \textbf{6$^{th}$} & 400 & 67.3 & 455 & 5.9 & 3.3 & 5.53 & 2.3 \\
         \hline
         \textbf{8$^{th}$} & 407 & 51.57 & 618 & 7.9 & 5.7 & 6.11 & 1.5 \\
         \hline
    \end{tabular}
    \vspace{0.5cm}
\caption{Summary of the obtained pulse parameters in our experiments. The first column denotes the pulse repetition-rate, which was either the 2nd, 4th, 6th or 8th harmonic of the fundamental repetition rate ($f_1=189.5$MHz). TBP denotes the time-bandwidth product of the pulses.}
\label{tab:parameters}
\end{table*}

\begin{figure}[ht]
    \centering
    \includegraphics[width=0.6\textwidth]{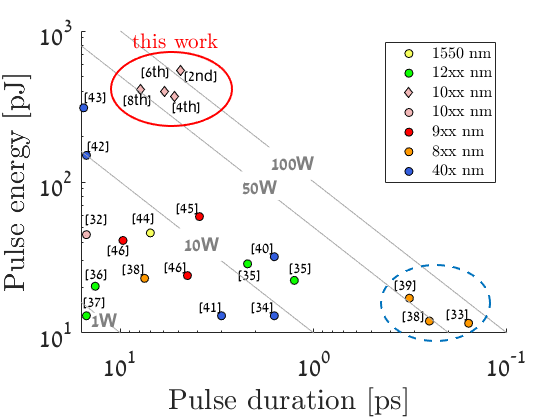}
    \caption{Comparison of our results (circled in red) to previously published results (diode edge emitter oscillator with an external cavity and no additional amplification stages) \cite{table1,table3,table4,table5,table6,table7,table8,table9,table12,table13,table14,table15,table18,table19,table20}. The axes indicate pulse-duration and pulse energy, while the straight lines indicate constant peak power.}
    \label{fig:comparisons}
\end{figure}

Figure \ref{fig:comparisons} provides a comparison of the obtained pulse performance to previously published systems in the literature. As evident, our pulse energies surpass previously published work by 1-2 orders of magnitude, while our peak power is 1-2 orders of magnitude higher than all ps-range sources, comparable only to fs-range oscillators (marked by a blue dashed circle on the figure) that have much lower pulse energy. Note that the pulse duration of our laser was far from minimal and not transform limited (with time-bandwidth product of 5.7 for the lowest harmonic) since our laser configuration did not include any measure for dispersion compensation, whereas the works circled with dashed blue in the figure did. Hopefully, proper dispersion compensation can push our pulse duration down as well, allowing to achieve higher peak powers at similar pulse energies. 

\begin{figure}[ht]
    \centering
    \includegraphics[width = 0.9\textwidth]{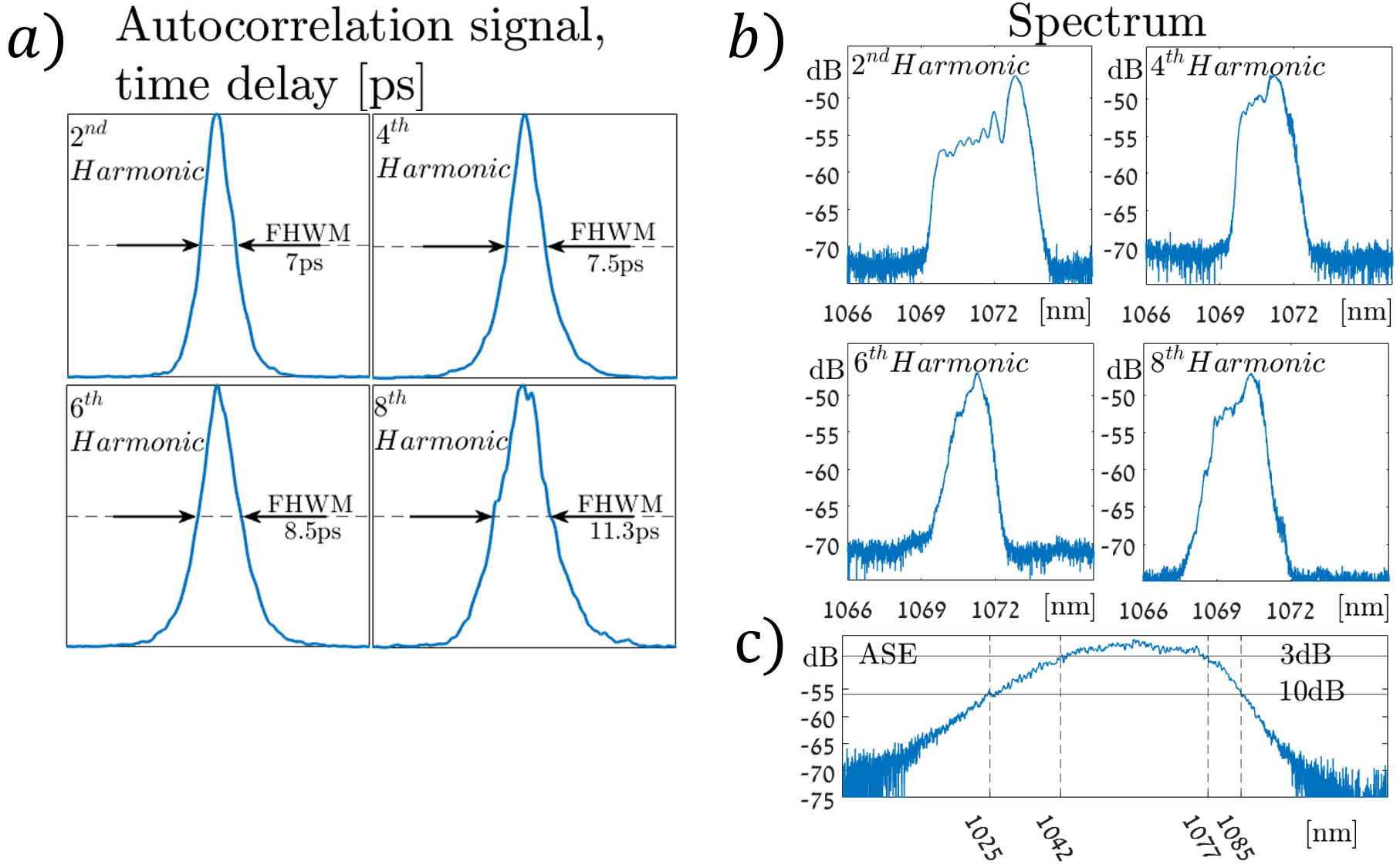}
    \caption{\textbf{Time and Frequency characterization of the pulses: (a)} Temporal intensity auto-correlation traces of the pulses for the various repetition rate harmonics: The pulse-lengths of table \ref{tab:parameters} are estimated assuming a Gaussian pulse, which yield a factor of 0.7 between the optical pulse and the auto-correlation trace. \textbf{(b)} The measured spectra of the pulses for different harmonics, which together with the auto-correlation measurements allows to calculate the time-bandwidth product of the pulses (TBP). \textbf{(c)} The spectrum of the amplified spontaneous emission, indicating the gain profile of the diode.}
    \label{fig:autocorrelation}
\end{figure}

Figure \ref{fig:autocorrelation} provides details of the measured pulses in both time and optical frequency. Figure \ref{fig:autocorrelation}a shows the auto-correlation traces of the different rep-rate harmonics and Figure \ref{fig:autocorrelation}b shows the optical spectrum of each harmonics, as well as the spontaneous-emission spectrum Figure \ref{fig:autocorrelation}c (the spectral gain curve). Clearly, the pulses are not transform limited as summarized in the table \ref{tab:parameters} above. 

\begin{figure}[ht]
    \centering
    \includegraphics[width=0.6\textwidth]{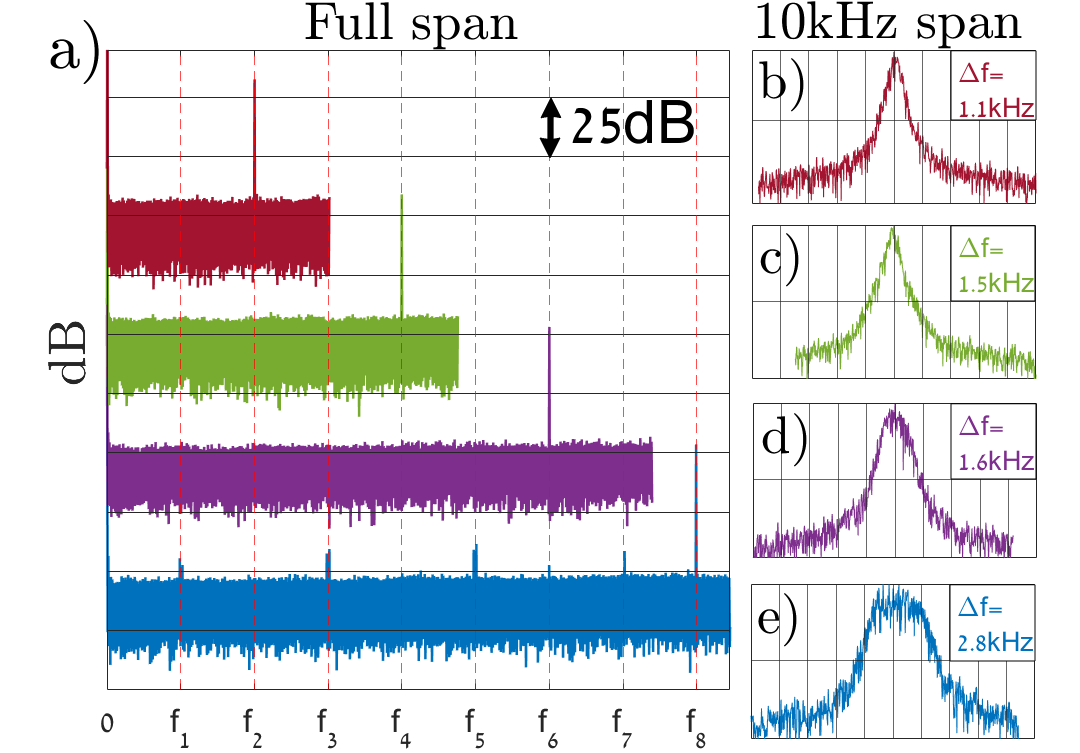}
    \caption{\textbf{Stability of the pulse train: RF spectrum} of the photo-detected pulses. \textbf{(a)} Full span of the spectrum, the measurement resolution bandwidth is 10kHz. \textbf{(b)-(e)} Narrow span ($10 \rm{kHz}$) around the n-th mode, with frequencies $f_{2}=379\rm{MHz}$, $f_{4}=758\rm{MHz}$, $f_{6}=1.137\rm{GHz}$, $f_{8}=1.516\rm{GHz}$ (resolution bandwidth of 10Hz).}
    \label{fig:rf_spectrum}
\end{figure}

The stability and purity of the generated pulse train is illustrated in Figure \ref{fig:rf_spectrum}. We analysed the RF spectrum of the pulse train, as measured on a fast photo-detector. A clean trace of the RF spectrum is obtained for 2nd,4th and 6th harmonics, showing only the oscillation rep-rate (and multiples) without any sub-harmonics or spurious frequencies (down to 50dB) below the, which indicates a clean, stable pulse train, whereas the 8th harmonics starts showing rep-rate instabilities. Detailed RF spectra of the rep-rate frequency are shown in figures \ref{fig:rf_spectrum}b-e, where the observed line-widths is  primarily due to the passive stability of the cavity length (which is not stabilized here at all). 

\begin{figure}[ht]
    \centering
    \includegraphics[width = 0.4\textwidth]{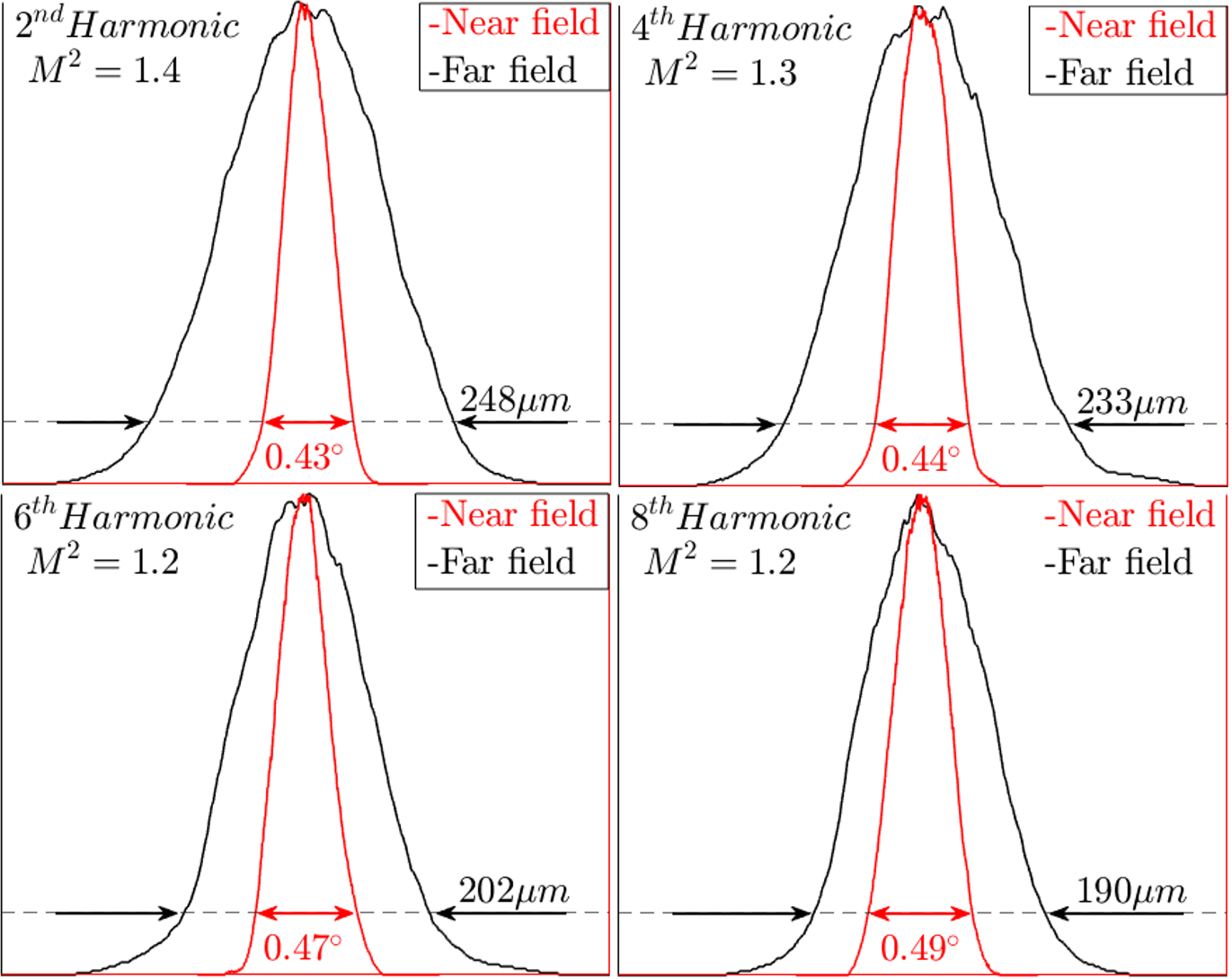}

    \caption{Beam intensity profile across the slow axis normalized by the peak intensity. The beam profiles were taken after filtering out the low intensity side lobes, which amount  to 12$\%$-14.5$\%$ of the total power. The width marked across the graphs is the width for which the distribution drops to 13.5$\%$ of the peak. 
    \textbf{(red)} Far field - the beam intensity profile as seen on the end-mirror. \textbf{(black)} Near field - beam intensity profile at the Fourier plane of the far field (that also corresponds to the field at the diode's facet). The calculated beam quality is varied between $M^2\!=\!1.2-1.4$.}
    \label{fig:M2_performance}
\end{figure}


Finally, our high-energy pulses are achieved with high spatial coherence, as illustrated in Fig. \ref{fig:M2_performance}. Specifically, $M^2=1.2-1.3$ is achieved for all the harmonics, indicating a near-ideal spatial mode with high brightness and focusability.

\section{Conclusions}
\label{sec:conclusions}
 
In conclusion, we present an external cavity configuration that implements passive mode-locking in a broad-area dual-section diode - a gain section with forward current and an integral saturable absorver section with a voltage controlled saturation level. The diode chip is cut at an angle to mitigate self-lasing at the diode facets. Our results demonstrate ps-range pulses with record performance in terms of pulse energy and peak power. The simplicity of the external cavity configuration that produces such high-energy and high-brightness ultrafast pulses is attractive for a wide range of applications, including laser micromachining, precision metrology, and nonlinear optics. 

The Possibility to produce even shorter, sub-ps pulses by incorporation of dispersion control in the cavity is fascinating and will be the focus of future research. 


\bibliography{main}
\end{document}

%% file: preamble.tex
\usepackage{amsthm}
\usepackage{mathtools}
\usepackage{physics}
\usepackage{xcolor}
\usepackage{graphicx}
\usepackage[left=23mm,right=13mm,top=35mm,columnsep=15pt]{geometry} 
\usepackage{adjustbox}
\usepackage{placeins}
\usepackage[T1]{fontenc}
\usepackage{lipsum}
\usepackage{csquotes}